\documentclass[twocolumn,floats,floatfix,superscriptaddress,aps,pra]{revtex4-2}
\usepackage{subfigure}
\usepackage{psfrag,graphicx}
\usepackage{bm,color}
\usepackage{latexsym}
\usepackage{epstopdf}
\usepackage[english]{babel}
\usepackage{amsmath,amssymb,amsfonts}
\usepackage{appendix}
\usepackage{bbold}
\usepackage[dvipsnames]{xcolor}

\definecolor{mygrey}{gray}{0.35}
\definecolor{myblue}{rgb}{0.2,0.2,0.8}
\definecolor{myzard}{cmyk}{0,0,0.05,0}
\definecolor{mywhite}{rgb}{1,1,1}
\definecolor{mywhite}{rgb}{1,1,1}
\definecolor{myred}{rgb}{1,0.,0.3}

\def\be{ \begin{equation}}
\def\ee{ \end{equation}}
\def\bse{  \begin{subequations}}
\def\ese{  \end{subequations}}
\def\bea#1\ea{\begin{align}#1\end{align}}

\def\bi{\begin{itemize}}
\def\ei{\end{itemize}}

\def\bt{\begin{tabular}}
\def\et{\end{tabular}}
\def\ket#1{\vert #1 \rangle}

\def\half{\tfrac12}

\def\3half{\tfrac32}

\def\to{\rightarrow}

\def\H{\mathbf{H}}
\def\S{\mathbf{S}}

\begin{document}

\author{K. N. Zlatanov}
\affiliation{Center for Quantum Technologies, Department of Physics, Sofia University, James Bourchier 5 blvd., 1164 Sofia, Bulgaria}
\affiliation{Institute of Solid State Physics, Bulgarian Academy of Sciences, Tsarigradsko chauss\'ee 72, 1784 Sofia, Bulgaria}
\author{S. S. Ivanov}
\affiliation{Center for Quantum Technologies, Department of Physics, Sofia University, James Bourchier 5 blvd., 1164 Sofia, Bulgaria}
%\author{B. W. Shore}
%\affiliation{618 Escondido Circle, Livermore, California 94550, USA}
\author{N. V. Vitanov}
\affiliation{Center for Quantum Technologies, Department of Physics, Sofia University, James Bourchier 5 blvd., 1164 Sofia, Bulgaria}
%\title{Creating coherent superpositions of N quantum states using Morris-Shore transformation}
%\title{\Red Creating arbitrary coherent superpositions of N quantum states using time-dependent Morris-Shore transformation \Black}
%\title{Extension of the Morris-Shore transformation to time-dependent fields: Application to coherent superpositions of multiple quantum states}
\title{Robust composite Molmer-Sorensen gate}
\title{Composite M{\o}lmer-S{\o}rensen gate}
%\title{\Red Creating arbitrary coherent superpositions of qudit states by double-adiabatic time dependent Morris-Shore transformation  \Black}
\date{\today }

\begin{abstract}
The M{\o}lmer-S{\o}rensen (MS) gate is a two-qubit controlled-phase gate in ion traps that is highly valued due to its ability to preserve the motional state of the ions. 
However, its fidelity is obstructed by errors affecting the motion of the ions as well as the rotation of the qubits. 
In this work, we propose an amplitude-modulated composite MS gate which features high fidelity robust to gate timing, detuning and coupling errors and is also tolerant of a.c. Stark shifts and drifting detuning errors. 
\end{abstract}

%\pacs{32.80.Bx, 33.80.Be, 34.50.-s, 32.80.Qk}
\maketitle

%--------------------------------------------

\section{Introduction}
 
Ion traps are one of the leading platforms for quantum information~\cite{Colin2019, Pogorelov2021}, simulation~\cite{Monroe2021,Manovitz2020} and sensing~\cite{Ilias2022,Gilmore2021}. For the past two decades these devices have excelled in single and two-qubit gate fidelity. 
Single qubit gates have achieved error rates as low as $1.5\times10^{-6}$~\cite{Leu2023} and even $10^{-7}$~\cite{Smith2024},  while two-qubit gate errors are currently of the order of $10^{-4}$ ~\cite{Clark2021,Gaebler2016}. 
Although such fidelity is below the fault tolerance limit the need for further improvements in it is ever present, since error correction is in general a resource-heavy procedure.

Standard ion traps employ spin-motion coupling to transmit interactions and hence they are prone to heating. The latter has a detrimental effect on the gate's fidelity and requires periodic cooling procedure. 
Among the different two-qubit gates proposed, the M{\o}lmer-S{\o}rensen (MS) gate~\cite{Molmer1998,Molmer2000} is highly cherished for its ability to preserve the motional state of the ions and thus avoid heating. 
The preservation of the motional state is caused by the path interferences induced by bichromatic excitation on both sidebands.
This interference remains not only for pure Fock states but also for coherent states~\cite{Ruzic2022}. 
Experimentally it was first demonstrated more than 20 years ago~\cite{Liebfried2003}, and more recently implemented on a chip surface traps~\cite{Hahn2019}, which underscores its efficiency and universality. 
The employed mechanism has also been investigated in other platforms like cavity QED~\cite{Takahashi2017} as well as in neutral atoms~\cite{Mitra2020}. Possible applications in Rydberg ions can be considered for the mitigation of motional effects over the gate's fidelity.
% In its essence the MS gate is a two-qubit rotational gate on the basis of which a bunch of other gates can be realized (?). 

The MS gate, however, is susceptible to various errors stemming from imperfect laser settings, interactions with the environment, cross-talk and  miscalibration. They can be categorized as motional and rotational errors, based on which component of the propagator they affect, and most of the listed above can affect both components. 
A number of techniques for the suppression and mitigation of these errors have been developed and experimentally demonstrated. 
For example, multi-tone excitation has shown mitigation of errors caused by heating of the motional mode~\cite{Webb2018}, as well as resilience versus gate timing errors and errors in the frequency of the motional mode~\cite{Shapira2018}. 
Pulse-shaping demonstrated resistance to disturbances of the motional mode~\cite{Duwe2022}. 
Amplitude modulation proved robustness towards normal mode frequency fluctuations~\cite{Zaran2019}. Frequency modulation techniques have been developed against static offsets of the motional-mode frequencies~\cite{Kang2021,Kang2023}. Phase modulations have shown robustness towards laser amplitude and gate detuning errors~\cite{Milne2020}. 
Most of these techniques correct for a single type of errors, therefore the question of combining techniques in order to correct for multiple errors simultaneously arises naturally.

In this paper, we propose the combination of amplitude modulation and composite pulses as a promising strategy to suppress multiple errors at once.  Composite MS gates have been proposed previously to correct for detuning errors by employing Walsh functions~\cite{Hayes2012} or to correct for Rabi frequency fluctuations in~\cite{Green2015} and in~\cite{Ivanov2015}. The later is devised under the strong assumption that errors only occur in the rotational component of the gate. 
This assumption can hardly be met experimentally, since  the control parameters that are quite probable to be the source of the errors like the Rabi frequency, the detuning and the timing of the gate affect both components.   
We demonstrate below that when such errors are accounted for in the motional component of the gate, the composite MS sequence can actually worsen its fidelity compared to a standard MS gate.
 To mend this problem we combine the sequence with amplitude modulation. This combination of techniques has been proposed earlier in a different context~\cite{Zhu2006}, namely developing an arbitrary speed gate by employing other modes in the presence of multiple ions. The technique of ~\cite{Zhu2006} demonstrates robustness to detuning errors, while errors in the Rabi frequency are acknowledged and dismissed as having negligible contribution. This is a very central distinction, since in our work, the aim of the composite sequence is to eliminate the rotational error, specifically the one that comes from errors in the coupling, and the method by which we achieve this is phase jumps in the laser fields addressing the sidebands, contrary to the specific detuning setting in~\cite{Zhu2006}.

We show, however, that when combined with the proper amplitude modulation, composite MS gates can outperform not only the standard MS gate configuration but also the multi-tone excitation schemes. What we demonstrate is that an amplitude modulation that is robust versus detuning and timing errors can be combined with the rotational robustness of composite pulses to provie an outcome which is robust to both motional and rotational errors and this robustness is maintained even if there are multiple simultaneous error sources.

This paper is organized as follows, Section \ref{section:two} introduces the dynamics of the MS gate. Section \ref{section:three} discusses the combination of amplitude modulation and composite sequences. In Section \ref{section:four} we demonstrate the robustness to multiple error sources acting simultaneously, and we conclude in Section \ref{section:five}.

%Developments in chip integrated surface ion traps seems as a very promising scalability solution, in which information processing is segmented to initialization, preparation and interaction regions on the chip. 

%--------------------------------------------
\section{Dynamics of the MS gate}\label{section:two}

The bichromatic excitation of the first red and blue sidebands of two trapped ions is described by the Hamiltonian
\be
H = g \sum_{k}\sigma(\zeta_k^+)\left( a^\dagger e^{i \epsilon t - i \zeta_k^-} + a e^{-i \epsilon t + i \zeta_k^-} \right),\label{Eq:H_MS}
\ee
where $\zeta^{\pm}_{k}=\half(\zeta^b_k \pm \zeta^r_k )$ is composed of the blue and red sideband phases, $\epsilon$ is the photon-phonon detuning to the intermediate state, $g= i\Omega \eta/2$ is the effective coupling and 
\be
\sigma(\zeta_k^+)= \sigma_k^+ e^{- i \zeta_k^+}+\sigma_k^- e^{i \zeta_k^+}
\ee
are the spin operators. 
%The simplified linkage pattern this Hamiltonian creates is illustrated in Fig.~\ref{fig:1}. 

%%%%%%%%%%%%%%%%%%%%%%%%%%%%%%%%%%%%%%%%%%%%%%%%%%%%%%%%%%%%%%%%%%%%%%%%%%%%%%%%%%%%%%%%%%%%%%%%%%%%%%%%%%%%%%%%55
%\begin{figure}[tb]
%\bt{cc}
%\includegraphics[width=0.85\columnwidth]{fig1.eps}
%\et
%\caption{  (Color online)
%Excitation scheme of the Molmer-Sorensen gate.  A combination of red and blue side-band excitation on each ion create an interference excitation path which allows a decoupling between the motional and spin degrees of freedom. The laser fields must be slightly detuned by $\epsilon$ from the photon-phonon resonance.}
%\label{fig:1}
%\end{figure}
%%%%%%%%%%%%%%%%%%%%%%%%%%%%%%%%%%%%%%%%%%%%%%%%%%%%%%%%%%%%%%%%%%%%%%%%%%%%%%%%%%%%%%%%%%%%%%%%%%%%%%%%%%%%%%%%%%%%%%%%%%%%%%%%%%%%%%%%

The propagator associated with Eq.~\eqref{Eq:H_MS} can be found by calculating the first two terms (all others vanish) of the Magnus expansion,
\bse
%\begin{subequations}
\bea
&M_1(t) = -\frac{i}{\hbar} \int_0^t H(t_1) \, dt_1,\\
&M_2(t) = \frac{1}{2} \left( \frac{i}{\hbar} \right)^2 \int_0^t \int_0^{t_1} [H(t_1), H(t_2)] \, dt_2 \, dt_1,
\label{Eq-Magnus b}
\ea
\label{Eq-Magnus}
\ese%\end{subequations}
and reads 
\be
U_{MS}=D(\alpha)e^{i \theta(t)\sigma(\zeta_1^+)\sigma(\zeta_2^+)},\label{Eq.-Ums}
\ee
where the displacement operator is given as
\be
D(\alpha) = \exp \left\{ \sum_k\sigma(\zeta_k^+)\left[\alpha_k(t) a^\dagger - \alpha_k(t)^* a \right]\right\}.
\ee
The functions  
%, which are determined by Eqs.~\eqref{Eq-Magnus}, and in more general form read
\bse
\bea
&\alpha_k(T_1,T_2) = - i \int_{T_1}^{T_2} g(\tau)\exp\left(i \epsilon \tau - i \zeta_k^{-}\right) \, d\tau , \\
&\theta(T_1,T_2) = 2 \int_{T_{1}}^{T_2}\int_{T_1}^{\tau_2}
   g(\tau_1) g(\tau_2) \sin  \left(\epsilon  (\tau_1-\tau_2\right))  d \tau_1   d \tau_2, 
\ea\label{Eq.-alpha-theta}
\ese
govern the evolution of the system for the time interval  $t\in [T_1,T_2].$ If $T_1=0$ we will use the short notation $\alpha(t),
\theta(t)$ where $t\in [0,T_2].$
For example, if we wish to generate  
$\ket{\Phi^{+i}}=\frac{1}{\sqrt{2}}(\ket{00}+i\ket{11}$  
we require $\alpha(\tau_g)=0$ and $\theta(\tau_g)=\pi/4$ at the gate time. 
This introduces a number of constraints between the control parameters.
For example, if standard constant coupling is employed, then
\bse
\bea
\epsilon= 4 g, \\
t \epsilon = 2 \pi. \label{Eq-teps}
\ea
\label{Eq-relations}
\ese
Errors stemming from a variety of sources can cause an incomplete trajectory in phase space or miss the gate time $\tau_g.$ For the vast majority of errors these two effects can happen simultaneously. 
This is a direct consequence of the model-specific parameter relations that arise, like the ones of Eqs.~\eqref{Eq-relations}. 
If an error occurs in one of the parameters it affects the relations to the others and thus lowers the fidelity. 
It is tempting to assume that errors emerge systematically in different parameters, say $t\to t(1+\delta)$ or $\epsilon\to\epsilon(1+\delta),$ where $\delta$ is the error.
%, have the same effect on the dynamics, since they affect the parameter relation in the same way. 
However, this is not the case %because they have a different effect on the gate's dynamics
and thus mitigation strategies aimed at fixing one can not always be used for the other, as we show in the next section. 
The protocol we propose aims to mitigate errors in the control parameters, $\epsilon$, $g$ and $\tau_g$, and thus to achieve a high fidelity gate even though the model-specific relations between the parameters may not hold.

%--------------------------------------------

One of the popular techniques for error mitigation is the multi-tone excitation \cite{Webb2018,Shapira2018,Lishman2020}, which relies on adding a number of laser beams shifted in frequency. 
The Hamiltonian of Eq.~\eqref{Eq:H_MS} is then transformed to \cite{Webb2018}
\be
H_{MT} =  \sum_{k}\sigma(\zeta_k^+)\sum_{j=1}^{n}g_j\left( a^\dagger e^{i j \epsilon t - i \zeta_k^-} + a e^{-i j \epsilon t + i \zeta_k^-} \right),\label{Eq-H-multitone}
\ee
where $j$ runs over the added tones and $g_j$ is given as $-0.1444\epsilon$ and $0.2888 \epsilon$ \cite{Haddadfarshi2016} for $j=1$ and $j=2$, respectively.
Usually only the first two tones are used, since experimentally adding enough power to the different tones is challenging \cite{Hensinger2024}. 
Due to the popularity of this technique we will use it as a reference point to which we will compare our protocol. For that purpose we will use the gate infidelity, defined as 
\be
I_g=1-F_g
\ee 
as a measure. Here the fidelity of the gate reads,   
\be
F_g=\frac{1}{K}\text{Tr}(U_{t}^{\dagger}U_{err}),\label{Eq:fidelity}
\ee
with K being the number of states, $U_{t}$ is the target propagator, while $U_{err}$ is the propagator in which an error  $s\to s(1+\delta)$  in one or multiple parameters has occurred. Of course in the latter case the different errors need not to be of equal magnitude. When a simulation of the full gate infidelity is inconvenient(see the subsection on a.c. Stark shift) we will use the infidelity for a state defined as
\be
I_s=1-\left| \langle \Psi_t\ket{\Psi_{err}}\right|^2.
\ee

\section{Composite gates with amplitude modulation}\label{section:three}

In this section we present the behavior of the composed modulated sequence versus different errors, acting alone. We combine these two techniques specifically because their individual set of constrains do not overlap. For example, we can not require the detuning to be such that the gate is insensitive to frequency and coupling errors since in general these constrains are different. For clarity, we discuss the amplitude modulation signals that are employed in Appendix~\ref{app:A}, for individual MS gates. Below we present the mechanism by which amplitude modulated composite gates are combined.

\paragraph*{}The idea of making a composite MS gate, that mimics composite pulses and corrects for rotational errors was first proposed in Ref.~\cite{Ivanov2015} and we refer the reader to this reference for a detailed derivation of the proposed sequences as well as any other details. Here we will only briefly introduce the procedure of generating robust composite MS gate. It uses the sequence 
\be
U^{(N)}(\theta) = F(\phi_{N+1}) U(\theta_N) F(\phi_N) \cdots U(\theta_0) F(\phi_0),\label{Eq.-full sequence}
\ee
where,
\bse
\be
U(\theta) = e^{i \theta \sigma_x \sigma_x},
\ee
\be
F(\phi) = e^{-i \phi \sigma_z}.
\ee
\ese
Such a sequence can be robust to rotational errors in the angle $\theta$ by a proper choice of the individual rotation angles $\theta_N$ and the phases of the $F(\phi_N)$ gates. 
The phase gates can be incorporated in the rotational gates by introducing 
\be
U_{\phi}(\theta) = e^{i \theta \sigma_x \sigma(\phi)}\label{Eq.Uc-theory},
\ee
where $\sigma(\phi) = \sigma_x \cos \phi + \sigma_y \sin \phi.$ 
Eq.~\eqref{Eq.-full sequence} then can be rewritten as
\be
U^{(N)}(\theta) = F(\phi_{N+1}) \prod_{k=0}^{N} U_{\phi_k}(\theta_k),
\label{Eq.-short sequence}
\ee
 and we outline that although Eq.(\ref{Eq.-full sequence}) and Eq.(\ref{Eq.-short sequence}) are mathematically equivalent, their physical realization is different. The former requires a single qubit gate implemented after each MS gate, while the latter achieves the effect of single-qubit rotations with phase-jumps of the bichromatic lasers. Thus a series of consecutive MS gates can realized without interruption and for all intended purposes we use the sequnce of Eq.(\ref{Eq.-short sequence}).  

In this manner broadband %and passband 
sequences can be generated if we make sure that
%\bse
\bea
&\left. \frac{\partial^l}{\partial \delta^l} \left[ U^{(N)}(\theta) - U(\theta) \right] \right|_{\delta=0} = 0
%&\text{and}\notag\\
%&\left. \frac{\partial^{l_1}}{\partial \delta_1^{l_1}} \left[ U^{(N)}(\theta) - U(\theta) \right] \right|_{\delta=0} = 0,\\
%&\left. \frac{\partial^{l_2}}{\partial \delta_2^{l_2}} \left[ U^{(N)}(\theta) - \mathbf{1} \right] \right|_{\delta=-1} = 0,
\ea
%\ese
is satisfied, where $l$ 
%and $l_i$ 
denotes different orders of correction. 

The propagator of Eq.~\eqref{Eq.Uc-theory} differs from Eq.~\eqref{Eq.-Ums} by the displacement operator. We can control the sign of the displacement $\alpha$ with the $\zeta_1^-=\zeta_2^-=\zeta^-$ phases. In this manner we can revert the displacement operator with two consecutive gates if we can ensure that $D(-\alpha)D(\alpha)=\mathbf{1}.$ This can happen if they have the same $\zeta_k^+$ phases, and also we choose the $\zeta^-$ phase of every second gate such that the relation
\be\label{Displacement closure}
D\left(\alpha(\zeta^-)\right)D(\beta)=e^{\frac{\alpha(\zeta^-)\beta^*-\alpha^*(\zeta^-)\beta}{2}}D\left(\alpha(\zeta^-)+\beta\right)=\mathbf{1}
\ee
will hold. This means that each logical gate in the sequence has to be represented by two physical gates that cancel each others displacement and rotate at half the angle as
\bea
U_{\phi}(\theta)_{logic}&= \underbrace{e^{i\frac{1}{2} \theta(t_2)\sigma_x\sigma(\zeta_2^+)}D(-\alpha)}_{2nd\quad gate}\underbrace{D(\alpha)e^{i \half \theta(t_1)\sigma_x \sigma(\zeta_2^+)}}_{1st\quad gate}\\ \notag &=e^{i \theta \sigma_x \sigma(\zeta_2^+)},
\ea
where $t_1$ and $t_2$ account for the proper time intervals of the gates. The phase $\zeta_2^+$ corresponds to $\phi,$ and is present in both the displacement and the rotational part of the propagator, thus we will always set $\zeta_1^+=0$ and control the rotation by the phase $\zeta_2^+.$

\begin{figure}[tb]
    \centering
    \begin{subfigure}
    \centering
        \includegraphics[width=.40\textwidth]{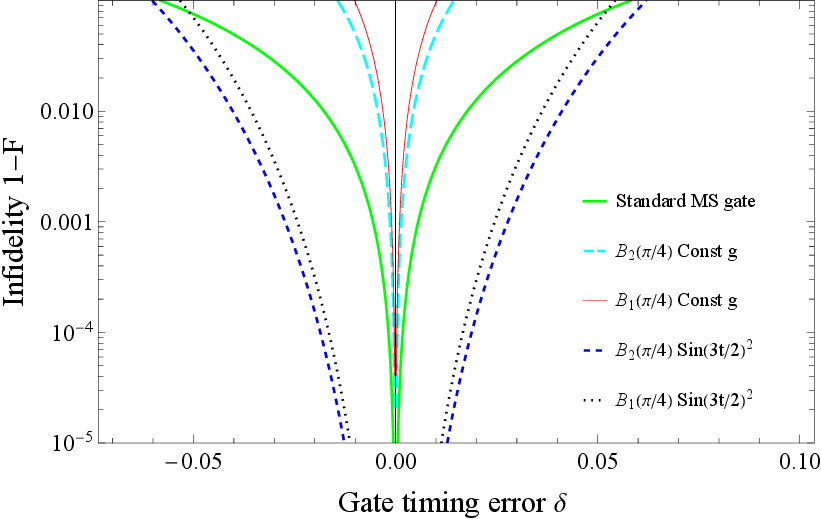}
        \label{fig:sub1}
    \end{subfigure}
    %\hfill % Adds space between subfigures
    \begin{subfigure}
        \centering
        \includegraphics[width=.43\textwidth]{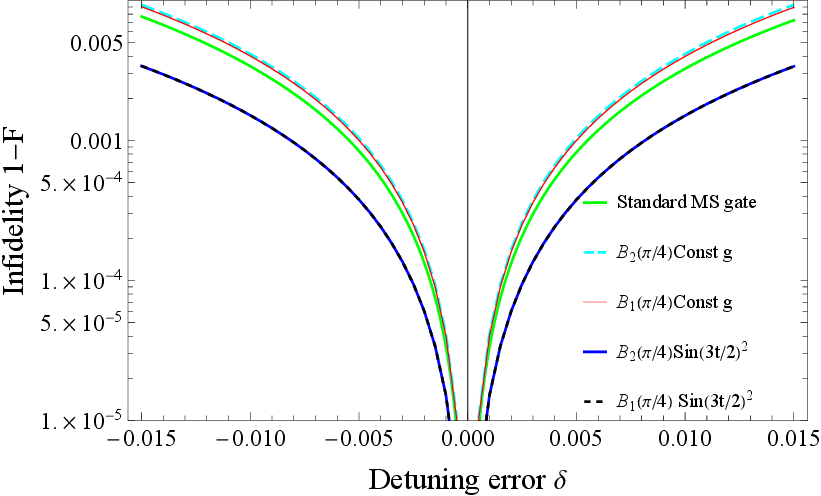}
        \label{fig:sub2}
    \end{subfigure}
   \caption{  (Color online)
Infidelity versus gate timing error(top) and detuning error(bottom) for $\mathbf{B_1}(\pi/4)$ and $\mathbf{B_2}(\pi/4)$ with constant signal (cyan and red lines) and for $\sin(3t/2)^2$ modulation (blue and black lines), compared to a standard MS gate. All sequences account for errors in the displacement operator as well. All sequences span an equally divided time interval of $2\pi,$ the detuning is $\epsilon/2\pi=1$ $[\tau_g^{-1}]$ and the couplings are set such that the required rotation for each gate in the sequence is achieved.}\label{fig:1}
\end{figure}

The shortest sequence we can use consists of three logical gates~\cite{Ivanov2015} and one single qubit phase gate,
\be
\mathbf{B}_1(\theta) =  F\left(-2\phi \right)\left( \frac{\pi}{2} \right)_{3\phi}  \left( \frac{\pi}{2} \right)_{\phi} (\theta)_0, \label{Eq.-B1}
\ee
where the sequence acts from right to left, that is the $(\theta)_0$ gate is implemented first.
This sequence however underperforms against gate timing errors (see Fig.~\ref{fig:1} top panel) for both constant and modulated signal. In our simulations of composite sequences we assume that timing errors will affect only the last physical gate in the sequence. The reason for this assumption is that generating a sequence of pulses with an acousto-optical modulator(AOM) is quite precise in the timing between individual pulses. Thus we assume that the timing error is caused by a shutter that opens or closes prematurely, and in this way cuts short the last physical gate or prolongs it. In this scenario, the error in the $\mathbf{B_1}(\theta)$ sequence will also affect the single qubit phase gate. Also, depending on the modulation signal, it might not preserve the plateau of the gate region, losing the advantage of amplitude modulation.  In order to avoid that we will use a permuted version, which performs better, namely
\be
\mathbf{B}_2(\theta) =  (\theta)_0  \left( \frac{\pi}{2} \right)_{\phi} \left( \frac{\pi}{2} \right)_{3\phi}  F\left(2\phi \right) . \label{Eq.-B2}
\ee
In this sequence the rotating gate comes last and this ensures that the timing error will not affect any other gate in the sequence and also that any robustness of the modulation will be inherited by the sequence.
%for broadband pulses which correct for 1st and 2nd order. And for passband sequence we will use
%\be
%\mathbf{P}_{1,1}(\theta) = (\pi)_{-\phi_1} (\pi)_{\phi_1} (\theta)_0. \label{Eq.-P1}
%\ee
Although longer sequences will compensate for higher-order errors, they can also increase the gate time. Even the sequence of Eq.~\eqref{Eq.-B2} can be longer by a factor of 2 to 3 than the single MS gate, depending on the employed modulation and the available power. 
%, based on the employed modulation. 
Hence we will limit ourselves only to short sequences like $\mathbf{B}_2(\theta)$. 
The phase gate in these sequences is on the second qubit and the phases are set to $\phi=\arccos(-\theta/\pi)$.

  %This means that for two consecutive time intervals the phases 
%The phase gates allow to be incorporated into the 
%where the individual propagator with the  is given as $
%U_{\phi}(\theta) = e^{i \theta \sigma_x \sigma_{\phi}}.$

%%%%%%%%%%%%%%%%%%%%%%%%%%%%%%%%%%%%%%%%%%%%%%%%%%%%%%%%%%%%%%%%%%%%%%%%%%%%%%%%%%%%%%%%%%%%%%%%%%%%%%%%%%%%%%%%55
%%%%%%%%%%%%%%%%%%%%%%%%%%%%%%%%%%%%%%%%%%%%%%%%%%%%%%%%%%%%%%%%%%%%%%%%%%%%%%%%%%%%%%%%%%%%%%%%%%%%%%%%%%%%%%%%55
%%%%%%%%%%%%%%%%%%%%%%%%%%%%%%%%%%%%%%%%%%%%%%%%%%%%%%%%%%%%%%%%%%%%%%%%%%%%%%%%%%%%%%%%%%%%%%%%%%%%%%%%%%%%%%%%%%%%%%%%%%%%%%%%%%%%%%%%
%%%%%%%%%%%%%%%%%%%%%%%%%%%%%%%%%%%%%%%%%%%%%%%%%%%%%%%%%%%%%%%%%%%%%%%%%%%%%%%%%%%%%%%%%%%%%%%%%%%%%%%%%%%%%%%%%%%%%
\begin{figure}[tb]
  \centering
  \bt{cc}
    % First row
    \includegraphics[width=42mm]{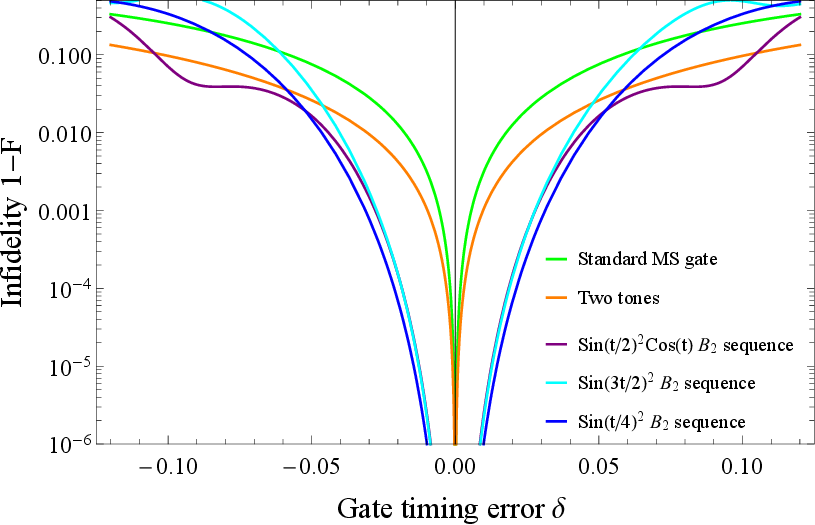}
    \llap{\parbox[b]{3.45 in}{(a)\\\rule{0ex}{0.85in}}}
    \hspace*{1mm} % Horizontal spacing between figures
    \includegraphics[width=42mm]{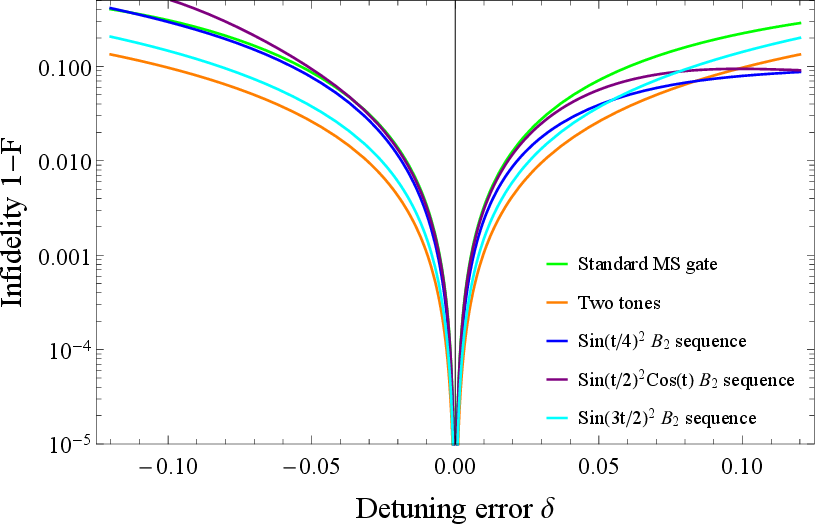}
    \llap{\parbox[b]{3.45 in}{(b)\\\rule{0ex}{0.85in}}}
    \\[5mm] % Vertical spacing between rows
    
    % Second row
    \includegraphics[width=42mm]{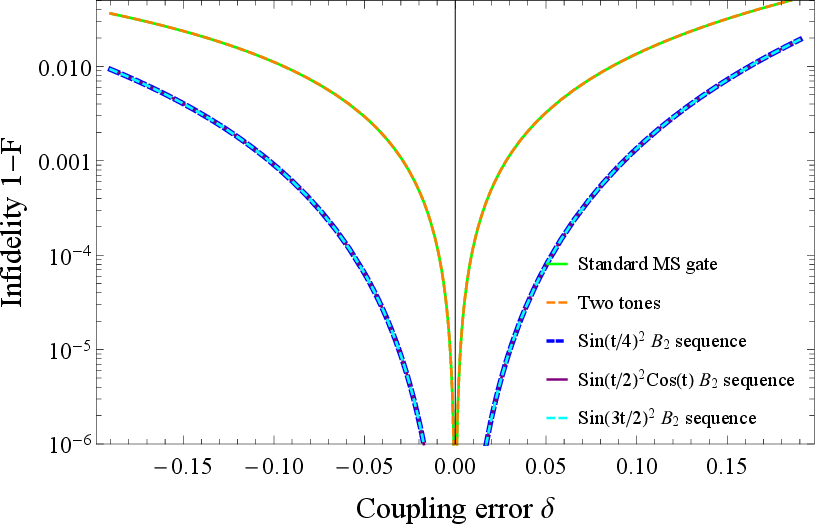}
    \llap{\parbox[b]{3.45 in}{(c)\\\rule{0ex}{0.92in}}}
    \hspace*{1mm} % Horizontal spacing between figures
    \includegraphics[width=42mm]{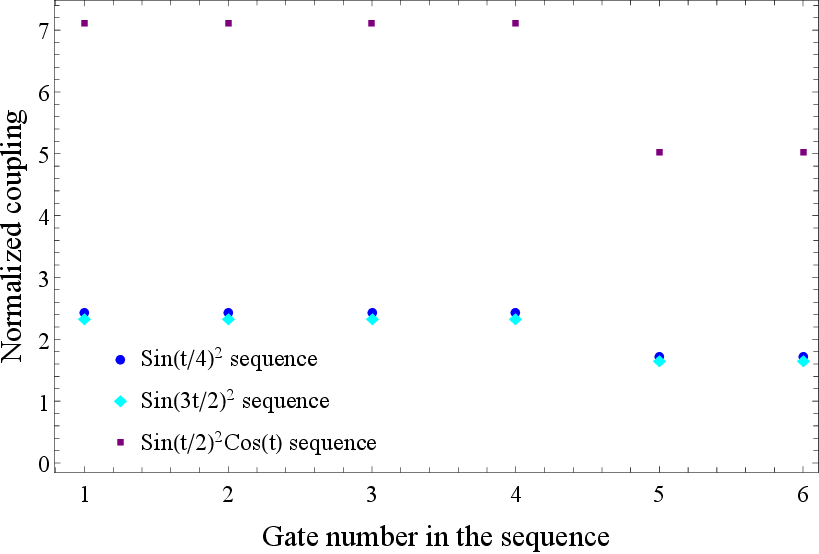}
    \llap{\parbox[b]{3.45 in}{(d)\\\rule{0ex}{0.92in}}}
  \et 
\caption{  (Color online)
Infidelity versus error from specific source. Frame a) shows infidelity vs gate timing errors, frame b) versus detuning error, and frame c) versus coupling error. The $\zeta^-$ phase that revert the displacement on every second gate in the sequence for the $\sin(t/4)^2,\sin(3t/2)^2,\sin(t/2)^2\cos(t)$ read  $\left\lbrace 0,-\pi+0.6207,0 \right\rbrace $ respectively. The detuning reads $\epsilon\to\left\lbrace 1,1,3\right\rbrace [\tau_g^{-1}],$ and the gate times are $\tau_g\to\left\lbrace 12\pi,2\pi,6\pi\right\rbrace.$ %\textbf{$2\pi$ zashto ne e 0?} 
Coupling strength of each gate in the sequence normalized to the coupling of a single MS gate of the specific modulation.
}\label{fig:2}
\end{figure}
%%%%%%%%%%%%%%%%%%%%%%%%%%%%%%%%%%%%%%%%%%%%%%%%%%%%%%%%%%%%%%%%%%%%%%%%%%%%%%%%%%%%%%%%%%%%%%%%%%%%%%%%%%%%%%%%55

The robustness to detuning and timing errors depends strongly on the Rabi modulation and is unaffected by the modulation for coupling errors.   
This is illustrated in Fig.~\ref{fig:1} for the sequences of Eq.~\eqref{Eq.-B1} and Eq.~\eqref{Eq.-B2}. 
We see that the sequences with constant coupling perform worse than the standard gate against gate timing errors(top frame) and versus detuning errors(bottom frame). Meanwhile the modulated sequences outperform the standard gate. This only shows that the amplitude modulation is crucial for implementation of sequences designed for robustness versus rotational errors. 
We tested three modulating signals $\sin(t/4)^2,\sin(t/2)^2\cos(t)$ and $\sin(3t/2)^2.$ In Fig.~\ref{fig:2} a) we see that all amplitude modulated sequences outperform the standard and the two-tone gate versus the gate timing error, an expected result, since the standard gate is not robust towards a timing error and  we use a two-tone gate which is optimized for detuning errors. 
In frame b), we see that the amplitude modulations are all better than the standard gate versus detuning error, but only the $\sin(3t/2)^2$ modulation performs comparably to the two-tone implementation of the gate. 
Now this is unexpected based on the individual performances of the modulations as we see in Fig.~\ref{fig:App} d), where it is worse than the two-tones, and only the sine-cosine modulation is performing comparably. 
%It is quite tempting to assume that the worse the modulation performs individually, the better it will perform in a sequence.
%However this is not really the case, since the $\sin(t/4)^2$ modulation, which was the worst performing individual modulation, is not the best performing in a %sequence, at least not against any type of error. In frame c) we see that the sine-cosine and the $\sin(3t/2)^2$ modulated sequences coincide and give the %best performance.
%This only comes to illustrate that the behaviour of the sequence is not that dependent on the individual trajectory in phase space. 
%TOZI PARAGRAPH E IZLISHEN, PONEJE SE OSNOWAWA NA STRANNI PREDPOLOJENIQ.
In frame c) we see that the standard and the two-tone implementations have the same resilience to coupling errors, while the performance of each modulation is the same. This is because the robustness towards coupling errors is due to the sequence itself, and is quite robust to all kinds of rotational errors. 
This demonstrates how by combining the two techniques robustness to multiple error sources can be achieved. Based on the given examples the  $\sin(3t/2)^2$ modulated sequence gives the best result in terms of multi-error robustness and is as long as the standard MS gate. The reasons for this is related to its phase space trajectory and also to its fast oscillation during the sequence time ($t\in \left[0,2\pi\right]$). 
This modulation has three bell-shaped "pulses" during the sequence time(see Fig.~\ref{fig:App} a) cyan line up to the $2\pi$ interval) and this allows each logical gate to be implemented over one of the three pulses. Physically, each pulse implements two gates. Since the phase space trajectory after the first half of each pulse is not naturally closed by the second half of each pulse, we have to implement a phase jump $\zeta_k^-=-\pi + 0.6207$ on every second half(gate) of each of the three pulses. This closes the phase space trajectory in accordance to Eq.(\ref{Displacement closure}). It might be the case that the second half of each pulse is naturally closing the phase space trajectory. This is the situation for the $\sin(t/4)^2$ and the $\sin(t/2)^2\cos(t)$ modulations, then there is no need of a phase jump, $\zeta_k^-=0.$    
The power costs per gate for each modulation is illustrated in Fig.~\ref{fig:2} d) where each gate of the sequences is normalized to the coupling strength that a single MS gate needs for the specific modulation. The $\sin(3t/2)^2$ has the lowest power requirements, which again supports our claim that it is very suitable modulation for composite MS gate. All presented modulations require larger power for each gate in the sequence than the minimal power required for the single modulated gate, which is to be expected since we are forcing the same rotation for a shorter time period (say $\pi/4$ over $\tau_g/6$). This power requirement can be relaxed if we simply allow a slower gate, that is, a longer time interval for each gate in the sequence. 
\section{Simultaneous error sources\label{section:four}}
In this section we show what happens if multiple error sources act simultaneously. We also expand the different error sources by adding a.c. Stark shift, and also drifting detuning that changes linearly during the gate. We focus primarily on detuning and coupling errors, since they are quite common. We see this case illustrated in Fig.~\ref{fig:3} a). Evidently, low infidelities can be maintained without any significant drop compared to the case of single error source. We only plot infidelities lower than $10^{-2}$ so each contour falls below this threshold and the red colored contours account for infidelities smaller than $10^{-3}.$ Due to the complexity introduced by the a.c. Stark shift, an analytic expression for the propagator was not tractable. Consequently, we computed state infidelities directly for the $\ket{\Phi^{+i}}=\frac{1}{\sqrt{2}}(\ket{00}+i\ket{11}$ state, without deriving the full propagator. 
\paragraph*{}

\subsection{A.C. Stark shift}
Designing a technique that is robust against a.c. Stark shift is quite challenging because it introduces an $\S_z$ operator in the Hamiltonian \cite{Garcia2022}
\be
\label{Stark_H}
\H(t) = \lambda(t)\, \S_z - g(t) \S_{\zeta_+}\left( a^\dagger e^{i \epsilon t - i \zeta_k^-} + a e^{-i \epsilon t+ i \zeta_k^-} \right),
\ee
with $\lambda(t)=\lambda g(t)^2,$ where we have introduced the collective spin operators $\S_{\zeta_+}=\sigma(\zeta_1^+=0)+\sigma(\zeta_2^+)$ and $\S_z=\sum_{k}\sigma_z$ in correspondence with Eq.(\ref{Eq:H_MS}).
Since the $\S_z$ operator does not commute with $\S_{\zeta_+},$ we can either use an approximation for the propagator keeping only the linear in $\lambda$ terms(meaning $M_3$ and $M_4$ terms in the Magnus expansion) as in \cite{Garcia2022} or simulate this numerically. In order to avoid the accumulation of errors due to the approximation through out the six pulses in the sequence we simulate the protocol numerically. We use a feed forward algorithm, solving the Schrodinger equation, with the Hamiltonian of Eq.(\ref{Stark_H}), where the output state of one time interval is fed as an initial condition to the next interval. Over the six gates numerical errors can accumulate on the order of the infidelity, thus we used 35 Fock states. Throughout this section we only give the results for the $g(t)=g\sin (3t/2)^2$ modulation, because it outperforms the other two. %The effect of the a.c. Stark shift is illustrated in Fig.~\ref{fig:3} b) and c). 
%%%%%%%%%%%%%%%%%%%%%%%%%%%%%%%%%%%%%%%%%%%%%%%%%%%%%%%%%%%%%%%%%%%%%%%%%%%%%%%%%%%%%%%%%%%%%%%%%%%%%%%%%%%%%%%%55
\begin{figure}[tb]
  \centering
  \bt{cc}
    % First row
    \includegraphics[width=42mm]{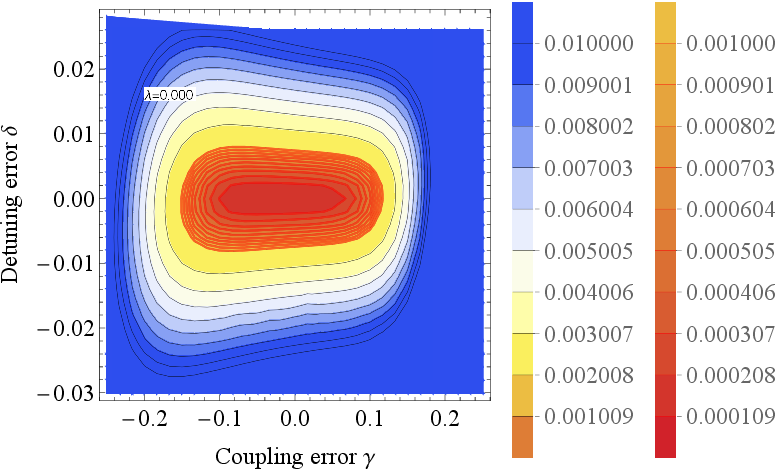}
    \llap{\parbox[b]{3.45 in}{(a)\\\rule{0ex}{0.85in}}}
    \hspace*{1mm} % Horizontal spacing between figures
    \includegraphics[width=42mm]{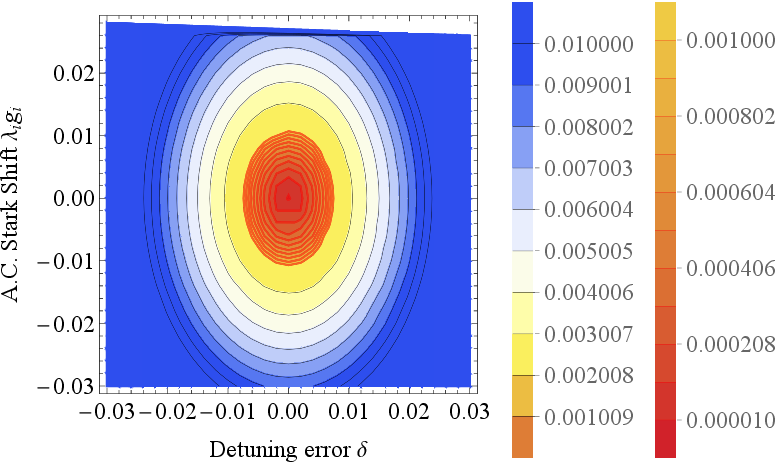}
    \llap{\parbox[b]{3.45 in}{(b)\\\rule{0ex}{0.85in}}}
    \\[5mm] % Vertical spacing between rows
    
    % Second row
    \includegraphics[width=42mm]{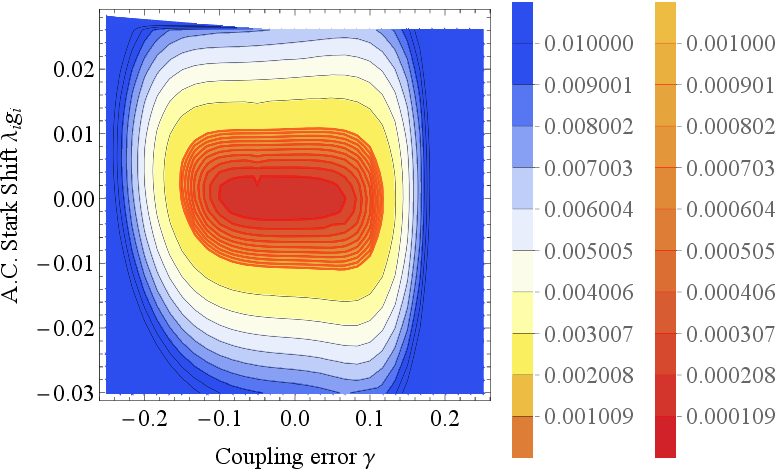}
    \llap{\parbox[b]{3.45 in}{(c)\\\rule{0ex}{0.85in}}}
    \hspace*{1mm} % Horizontal spacing between figures
    \includegraphics[width=42mm]{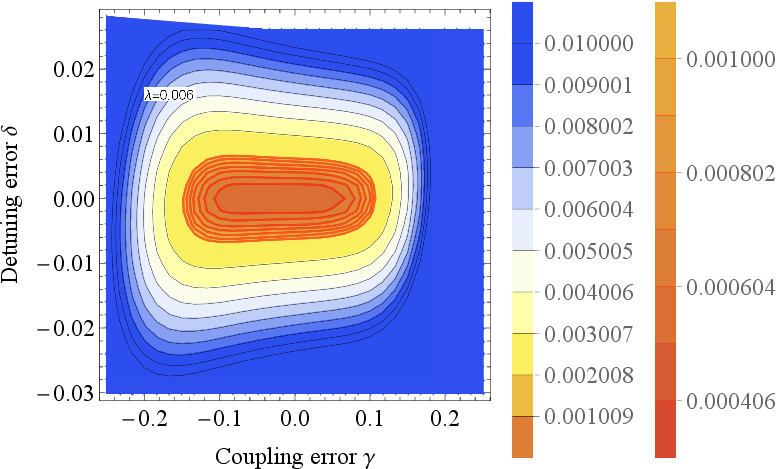}
    \llap{\parbox[b]{3.45 in}{(d)\\\rule{0ex}{0.85in}}}
  \et
  
  \caption{(Color online) Infidelity contour plots for multiple error sources acting simultaneously. In frame a) these are static detuning and coupling error, in b) a.c. Stark shift and static detuning error in c) a.c. Stark shift and coupling error and in d) static detuning and coupling error while there is a Stark shift with magnitude $\lambda=0.006g_i$, where $g_i$ is the coupling of the i-th pulse. In each frame the outer scale shows the infidelity in the red contours.} 
  \label{fig:3}
\end{figure}
%%%%%%%%%%%%%%%%%%%%%%%%%%%%%%%%%%%%%%%%%%%%%%%%%%%%%%%%%%%%%%%%%%%%%%%%%%%%%%%%%%%%%%%%%%%%%%%%%%%%%%%%%%%%%%%%%%%%%%%%%%%%%%%%%%%%%%%%
%%%%%%%%%%%%%%%%%%%%%%%%%%%%%%%%%%%%%%%%%%%%%%%%%%%%%%%%%%%%%%%%%%%%%%%%%%%%%%%%%%%%%%%%%%%%%%%%%%%%%%%%%%%%%%%%55
The key indicator of the effect the Stark shift has is the relative magnitude to the coupling $g.$  Each pulse in the sequence creates its own Stark shift so in the simulations the magnitude is taken as parts of the coupling for the specific pulse. Since the coupling depends also on the Lamb-Dicke parameter we expect the Stark shift to be on the order of $10^{-2}$ to $10^{-3}$ parts of the coupling. In any case this is the range to which our technique can absorb the effects of the Stark shift. This is illustrated in panels b) and c) of Fig.~\ref{fig:3} for Stark shift and detuning error and Stark shift and coupling error respectively. In frame d) we illustrate the robustness to detuning and coupling while a Stark shift of magnitude $\lambda=0.006g_i$ is present. We chose this value for the simulation because it falls within the region of $10^{-3}$ infidelity in the other frames. As can be seen it has an effect on the robustness region, but still a decent infidelities can be maintained.      
\subsection{Drifting detuning error}
\begin{figure}[tb]
    \centering
    \begin{subfigure}
    \centering
        \includegraphics[width=.50\textwidth]{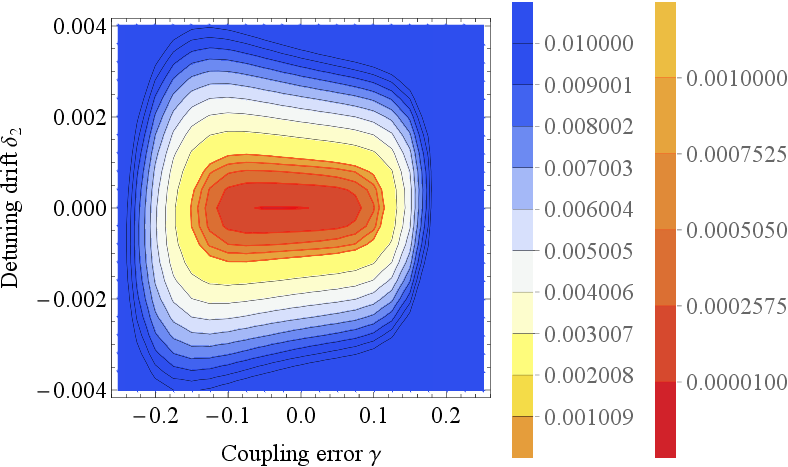}
    \end{subfigure}
    %\hfill % Adds space between subfigures
    \begin{subfigure}
        \centering
        \includegraphics[width=.50\textwidth]{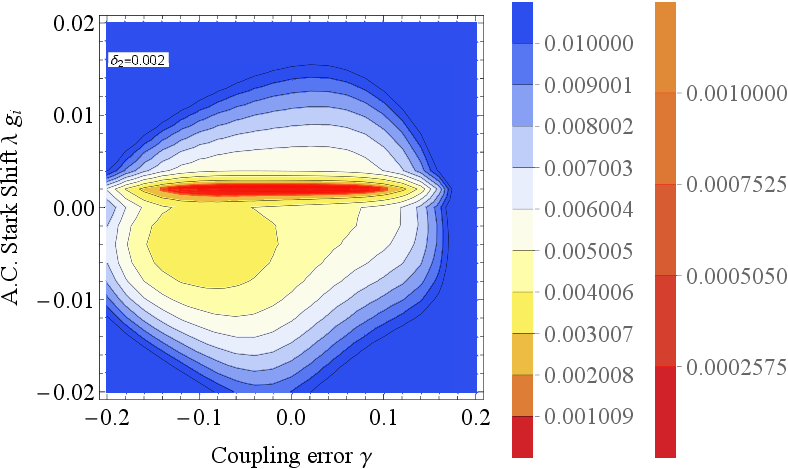}
    \end{subfigure}
    \caption{Contour plots for the infidelity versus drift rate and coupling error in the absence of Stark shift(top) and versus Stark shift and coupling error(bottom) for fixed drift rate $\delta_2 =0.002.$ The presence of multiple errors creates local minima besides the global one. The position of the minima can be mirror reflected by the sign of the drift.}
    \label{fig:4}
\end{figure}

The detuning does not necessarily needs to be static, but can rather be drifting during the gate, and presumably each gate in the sequence may have different detuning.  Here we assume a linear drift $\epsilon\to \epsilon(1+\delta_1+\delta_2 t),$ where $\delta_1$ is the static error and $\delta_2$ is the drift rate. Techniques that are specifically designed to handle drifting detuning can account for such rates that an error in the range 0.5 to 1.5 kHz is accumulated during a few tens of $\mu s$ gate times \cite{Vedaie2023,Leung2018}, however \cite{Li2022} reports drifts on the Hz scale in the time range of hours. In any case our technique can account for drift rates up to $\delta_2 =0.004[t^{-1}]$ as illustrated in Fig.~\ref{fig:4}. There are two effect that the static error has while the drift occurs. Regardless if its accumulating from pulse to pulse or if it remains the same, while the frequency is drifting, the first effect is to shift the minimum of the infidelity with direction depending on the sign of the static error. The second effect is to increase the infidelity, eating away the error budget for the drift. For this reasons we are showing a simulation with a drifting error only, that is $\delta_1=0$.  
%%%%%%%%%%%%%%%%%%%%%%%%%%%%%%%%%%%%%%%%%%%%%%%%%%%%%%%%%%%%%%%%%%%%%%%%%%%%%%%%%%%%%%%%%%%%%%%%%%%%%%%%%%%%%%%%55
%%%%%%%%%%%%%%%%%%%%%%%%%%%%%%%%%%%%%%%%%%%%%%%%%%%%%%%%%%%%%%%%%%%%%%%%%%%%%%%%%%%%%%%%%%%%%%%%%%%%%%%%%%%%%%%%%%%%%%%%%%%%%%%%%%%%%%%%
From the top frame of Fig.~\ref{fig:4} we see that in the absence of Stark shift the behaviour of the infidelity is not fundamentally different from what we had for a static detuning error in Fig.~\ref{fig:3}a). However allowing a Stark shift makes a dramatic change as we see in the bottom frame, where we have the infidelity for Stark shift and coupling error at fixed drifting rate $\delta_2=0.002[t^{-1}],$ for which the infidelity is in the $10^{-3}$ region as indicated by the top frame. Since the a.c. Stark shift changes the energy levels of the internal states dynamically, the effect of the linear detuning drift can be enhanced or diminished. For example in the bottom frame, note that in the absence of a Stark shift, the infidelity remains at the $10^{-3}$ level, while with a positive shift it decreases to the $10^{-4}$ level. Similar effect can be achieved by a negative coupling error with a lesser extent, thus a local minimum can form. What we believe to be the physical explanation for this effect is that when more than two error sources are present, they can negate or enhance each others effect by creating a rotation and displacement in opposite directions. In such scenario a local minimum forms and the global minimum is shifted towards a certain value of the error source. Also the position of the minima can be reflected by changing the signs of the error sources.  

\subsection{ Application in ${}^{171}\text{Yb}^+$}
\begin{figure}[tb]
    \centering
    \begin{subfigure}
    \centering
        \includegraphics[width=.45\textwidth]{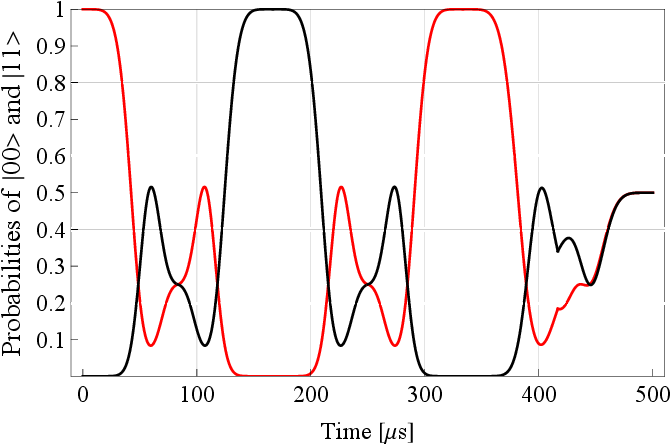}
    \end{subfigure}
    %\hfill % Adds space between subfigures
    \begin{subfigure}
        \centering
        \includegraphics[width=.45\textwidth]{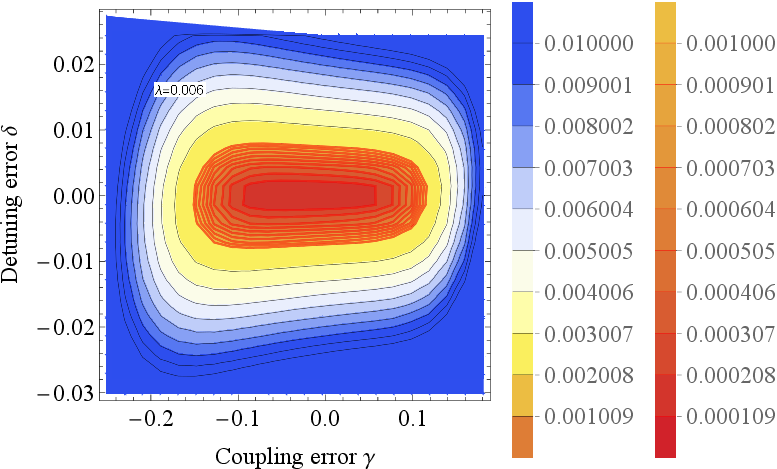}
    \end{subfigure}
    \caption{The $ \sin(3t/2)^2$ modulated $B_2$ sequence for ${}^{171}\text{Yb}^+$ for a system initialized in the $\ket{00}$ state. Top: Populations of $\ket{00}$(red curve) and $\ket{11}$(black curve) during the sequence of 500 $\mu s,$ each pulse is detuned at  $\epsilon/2\pi=1 / \tau_g.$ Bottom: Contour plots for the infidelity versus static detuning and coupling error in the presence of a.c. Stark shift of $\lambda_i=0.006 g_i.$ Scales indicate parts of the respective quantity e.g. coupling error of $\gamma= 0.1$ indicates an error of  $0.1g_i.$}
    \label{fig:5}
\end{figure}

We now illustrate the application of our technique in a real system, namely ${}^{171}\text{Yb}^+.$ A common choice for the qubit is in the ${}^{2}\text{S}_{1/2}$ manifold, encoded in the $\ket{F=0,m_F=0}$ and $\ket{F=0,m_F=1}$ states with a hyperfine splitting of $\omega_{HF}=12.642812$ GHz. This transition is addressed with a Raman beam pair via the ${}^2P_{1/2}$ manifold. We assume a Lamb-Dicke parameter of $\eta=0.065$. We simulate a gate time of $83.33\mu s,$ for each gate in the sequence, thus for the six physical gates the total time of the sequences is 500 $\mu s,$ and the detuning for each pulse is $\epsilon/2\pi=1 / \tau_g= 2$ kHz. The Rabi frequencies needed for this configuration are $\Omega/2\pi= g/(\pi\eta)=214.528 $ kHz for the first four gates and 151.694 kHz for the last two. Each pulse in the sequence generates its own a.c. Stark shift. The tolerable level for which the sequence can maintain infidelity lower that $10^{-3}$ is $\lambda_{1-4}=262.84$ Hz for the first four pulses and $\lambda_{5-6}=185.85$ Hz for the last two. Beyond this limit the Stark shift needs to be compensated with an external field  or alternatively with a combination of wave plates \cite{Tu2025}. We assume that the pulses are generated by an AOM as in \cite{Blumel2021}, and that it can also implement the phase jumps similar to \cite{Green2015}. We illustrate the dynamics and the infidelity in Fig.~\ref{fig:5}. As evident from the dynamics, the time robustness is maintained both in the beginning and in the end of the sequence and for that reason its irrelevant when timing error occurs. From the bottom frame we see that the tolerance(infidelity$<10^{-3}$) for coupling errors in the range of $15\%$(6.57 kHz) to $11\%$(4.81 kHz), depending on the sign of the error and similarly for detuning errors of $0.8\%$(160 Hz) in each direction. 
%%%%%%%%%%%%%%%%%%%%%%%%%%%%%%%%%%%%%%%%%%%%%%%%%%%%%%%%%%%%%%%%%%%%%%%%%%%%%%%%%%%%%%%%%%%%%%%%%%%%%%%%%%%%%%%%55

\section{Discussion and conclusion}\label{section:five}
In this work we presented the combined technique of amplitude modulation and composite gates to produce a Molmer-Sorensen gate which is robust to timing errors, rotational errors and detuning errors and can also withstand certain levels of a.c. Stark shifts and drifting detuning. 
To our knowledge, the sine-cosine amplitude modulation is new and individually it performs quite well against both timing and detuning errors. 
The permuted sequence of Eq.~\eqref{Eq.-B2} that we propose negates the effects of errors propagation from the two-qubit part of the sequence to the single-qubit gates in it and maintains the robustness of the amplitude modulation versus gate timing errors. 
The $\sin(3t/2)^2$ modulated sequence outperforms the two-tone excitation technique versus all types of errors except for an error in the detuning, although its performance there is quite similar. It also surpassed the sine-cosine modulation, which performs better individually. This is due to errors in the displacement operator. We came to this conclusion by comparing the behaviour of the sequences under the assumption of perfect displacement operators which are errorless then the behaviour is quite similar to the single gate case where the $\sin^2(t/2)\cos(t)$ signal is the most robust one. Individually, the $\sin^2(t/2)\cos(t)$ modulation outperforms the rest of the signals if its trajectory is allowed to complete a full traversing of the phase space and end back in the origin. If we interrupt it during the interaction, in order to speed-up the procedure the over-all error increases. This is due to the $\cos(t)$ part of the modulation, since it is responsible for the wings of the pulse, which generate the small arcs in the beginning and by the end of the evolution in phase space(see Fig.~\ref{fig:App} b)). During the evolution the curve acquires a phase which at specific intervals rephases the trajectory in a similar way to dynamical decoupling. For the $\sin^2(t/2)\cos(t)$ modulation this effect is most profound for the wings of the pulse, therefore disrupting the trajectory loses the strength of the technique. Based on that we can conclude that the best performing modulation individually is not necessarily the best performing modulation in a sequence. Furthermore, the $\sin(t/4)^2$ modulation's trajectory is very similar to the two-tone's trajectory, yet their performances, both individually and in a sequence, differ strongly. Thus a clear indication for the performance of a MS gate, based only on its phase-space trajectory can not be derived.    
There is a clear trade-off between time, power and robustness which is an expected effect, also observed  in single qubit gates \cite{Ballance2016}. 
In general, the power costs of a sequence with modulated Rabi frequency are a bit below 2.5 times the cost of a single MS gate with the same modulation. 
Different modulations allow for a variety of gate times ranging from the minimal gate time for a single MS gate to a few multiples of it.
However, choosing a proper signal for modulation in a composite gate has to be in accordance with the power capabilities of the implementing platform and its individual tolerances to errors of different nature. 

\acknowledgements
 This research is supported by the Bulgarian national plan for recovery and resilience, Contract No. BG-RRP-2.004-0008-C01 (SUMMIT), Project No. 3.1.4 and by the European Union’s Horizon Europe research and innovation program under Grant Agreement No. 101046968 (BRISQ). We are greatful to the following projects VU-F-205/2006,DO-02-136/2008,DO-02-167/2008,DO 02-90/2008 for granting access to the PHYSON cluster. 
\appendix
\section{Amplitude Modulation}\label{app:A}
Here we introduce the technique of amplitude modulation for two-qubit gates. 
Ultimately our goal is to combine it with composite pulses keep their benefits while diminishing their flaws.
We will employ amplitude modulation for its ability to mitigate timing and detuning errors and we will also join together multiple gates, generating a sequence, mimicking composite pulses techniques, to eliminate rotational errors.   %Accounting for the motional errors however leads to a sensing mechanism, whose sensitivity grows with the length of the sequence.
%\subsection{Amplitude modulation}
In order to find a proper time-dependent Rabi frequency which will be able to produce the MS gate, there are a couple of assumptions we can make based on Eqs.~\eqref{Eq-Magnus}. 
First, we demand that the $M_1(t)$ term evaluates to 0 at the gate time which will ensure a closed trajectory in the phase space. 
Second, due to the form of the $M_2(t)$ term, the Rabi frequency must have a good auto-correlation in order to achieve a strong excitation with minimal amount of power. 
Both of these assumptions have to be satisfied simultaneously. 
A natural candidate functions that satisfy these requirements are the trigonometric functions. 
Using a sine function provides a smooth-start~\cite{Sutherland2019}, which minimizes the coupling to the carrier and also has proven to have a low power requirement~\cite{Duwe2022}. 
In order to find a larger family of functions, we can combine it with a cosine of a certain degree, as long as the smooth start is ensured. 
In its most general form %~\cite{Zaran2019}% 
the coupling of Eq.~\eqref{Eq:H_MS} then reads
\be
g(t,m,l,n,p)=A \sin(m t)^l\cos (n t)^p.
\ee
Such amplitude modulation provides analytical expressions for $\alpha(t)$ and $\theta(t),$ for all different sets of parameters $m,l,n,p.$ They are too cumbersome to be presented here so instead we demonstrate the dynamics of a few of these modulations shown Fig.~\ref{fig:App}. We present their analytical form  used for this illustration in the Appendix~\ref{app:B}. 
These modulations generate a variety of trajectories, illustrated in Fig.~\ref{fig:App}(b), some quite similar and others quite different. The relation between the phase space trajectory and the robustness versus specific parameter is not fully understood. There are empirical observations which state that on average the trajectory must be as close to the center as possible. A strict mathematical derivation describing the relation between robustness and trajectory is yet to be derived. We note that there are attempts to achieve this with an interesting algebraic approach ~\cite{Duwe2022}.
%, which we find a bit too restrictive.
We now turn our attention to the selected modulations and point out that since the parameter space is quite rich the following discussion is not aimed at drawing the borders of the best performing parameter region, but rather to draw quantitative conclusions based on the given examples. 
For example, the sine-squared shape modulation is in general quite good for gate timing errors, as it transforms the point of entanglement into a plateau. Changing the frequency of the sine also seems to impact the infidelity, as can be seen from Fig.~\ref{fig:App}(c). 
%Such a frequency change, however, might increase the gate time, e.g. the $\sin(t/4)^2$ shape gives the slowest operation, yet it does not outperform the $\sin(t/2)^2\cos(t)$ modulation for timing errors. 
When we look at the performance against detuning errors, shown in Fig.~\ref{fig:App}(d), we see that the $\sin(t/2)^2\cos(t)$ outperforms the other sine shapes. Note the asymmetry in Fig.~\ref{fig:App}(d), which is due to a resonance. In this example of amplitude modulation a larger detuning is required so that a closed trajectory can be formed and consequently a stronger coupling is needed. Thus a clear trade-off emerges between gate-time, power and robustness to multiple parameters, that one has to keep in mind, when choosing a specific modulation signal with respect to the experimental platform at hand. 
%%%%%%%%%%%%%%%%%%%%%%%%%%%%%%%%%%%%%%%%%%%%%%%%%%%%%%%%%%%%%%%%%%%%%%%%%%%%%%%%%%%%%%%%%%%%%%%%%%%%%%%%%%%%%%%%%%5
\begin{figure}[tb!]
  \centering
  \bt{cc}
    % First row
    \includegraphics[width=52mm]{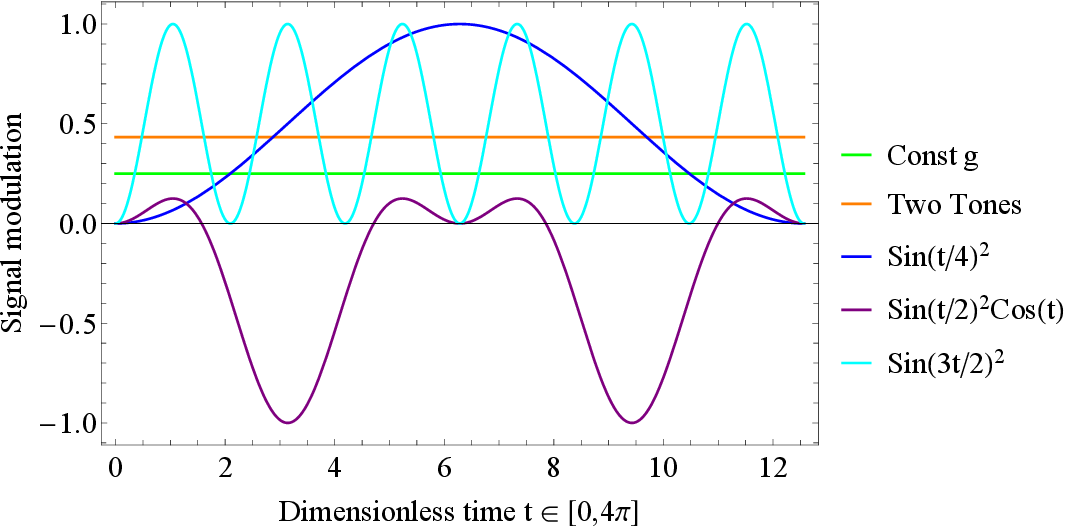}
    \llap{\parbox[b]{4.15 in}{(a)\\\rule{0ex}{0.85in}}}
    \hspace*{1mm} % Horizontal spacing between figures
    \includegraphics[width=33mm]{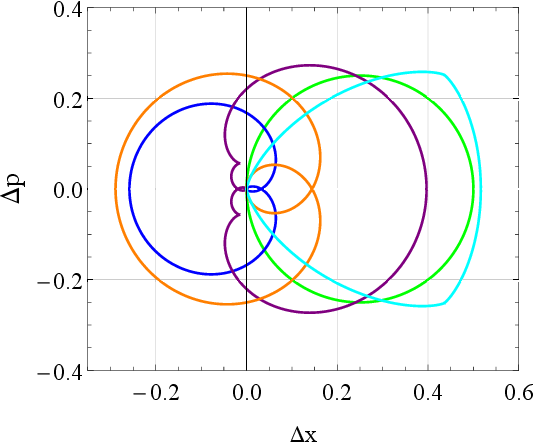}
    \llap{\parbox[b]{2.85 in}{(b)\\\rule{0ex}{0.85in}}}
    \\[5mm] % Vertical spacing between rows
    
    % Second row
    \includegraphics[width=42mm]{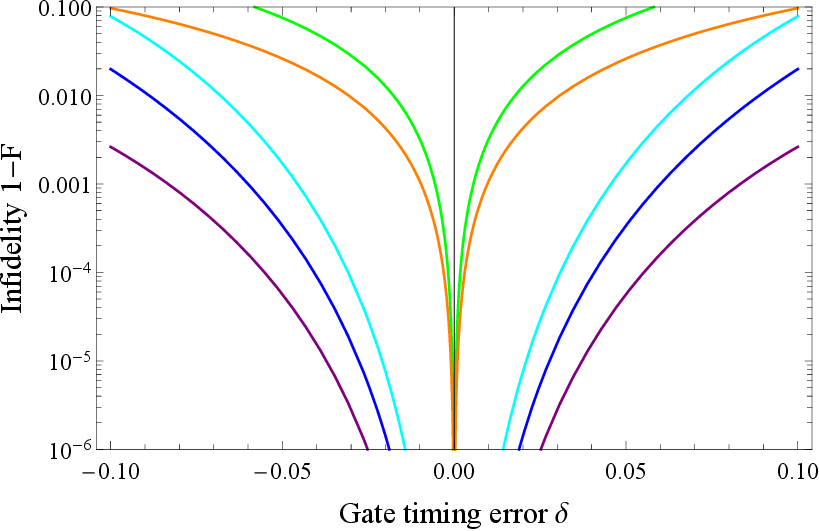}
    \llap{\parbox[b]{3.45 in}{(c)\\\rule{0ex}{0.85in}}}
    \hspace*{1mm} % Horizontal spacing between figures
    \includegraphics[width=42mm]{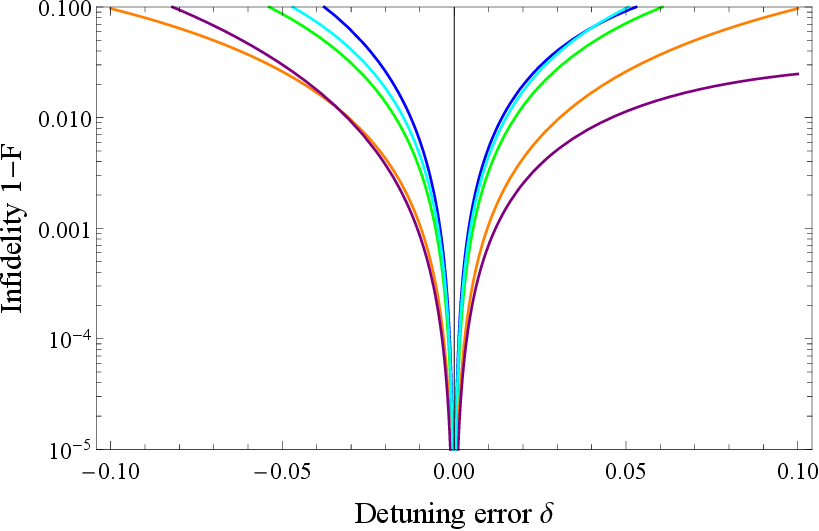}
    \llap{\parbox[b]{3.45 in}{(d)\\\rule{0ex}{0.85in}}}
  \et 
\caption{  (Color online)
MS gate under sine amplitude modulation. Frame a) shows the used signals for amplitude modulation and frame b) illustrates the phase space trajectories. Frames c) and d) depict the infidelity for gate timing and detuning errors respectively. The simulations were carried on the first possible trajectory for integer detuning, which for the order given in frame a) reads $\epsilon \to \left\lbrace 1,1,1,3,1\right\rbrace$ in units of $\tau_g^{-1}$  with the corresponding gate times and couplings given as $\tau_g \to \left\lbrace 2\pi, 2\pi, 4\pi, 2\pi, 2\pi \right\rbrace$ and $g \to \left\lbrace 0.25, 0.433,0.2733,0.85,0.516 \right\rbrace$ in corresponding units of frequency. All simulations are done with 14 motional states.
}\label{fig:App}
\end{figure}

%%%%%%%%%%%%%%%%%%%%%%%%%%%%%%%%%%%%%%%%%%%%%%%%%%%%%%%%%%%%%%%%%%%%%%%%%%%%%%%%%%%%%%%%%%%%%%%%%%%%%%%%%%%%%%%%55
%However, combining the sine with a cosine modulation, outperforms all pure-sine shapes, for gate timing errors, but also behaves just as good, if not better than the multi-tone excitation for detuning errors. 
In our simulations we use only two-tone excitation optimized for detuning errors, and for that reason it gains infidelity versus gate timing errors. We also point, however, that it can be optimized for gate timing errors as in Ref.~\cite{Shapira2018}. We see that for detuning errors the sine-cosine modulation can underperform or outperform the multi-tone excitation depending on the sign of the detuning error. In any case it is more advantageous to use such amplitude modulation since it also provides gate timing robustness. 
We will use it in a composite sequence in order to generate a gate that is robust against any type of error. As well as the other two signals, namely the $\sin(t/4)^2$ and the $\sin(3t/2)^2$ modulations. Thus we attempt to draw any conclusion about their performance based on their phase-space trajectory, as the former provides quite a different trajectory than the multi-tone excitation, while the later has its trajectory quite similar. In terms of fidelity the former performance is somewhat in the middle of all other shapes, while the later has the worst performing infidelities for gate timing errors. It is worthwhile to point out that these two modulations behave even worse than the standard gate versus detuning errors, and it is crucial to clarify that this behavior is due to the phase-space trajectory, which we can also change by the detuning,  rather than the modulation itself.  
\section{Analytical expressions for $\alpha$ and $\theta$}\label{app:B}
\begin{flushleft}
Here we provide analytical expressions for the functions of Eqs.(\ref{Eq.-alpha-theta}). We only show them for a time interval in which a rotation of $\pi/4$ is achieved and we have set all phases to 0 for simplicity.  Starting with the $\sin(t/4)^2$ modulation for $t \in [0,4\pi]$ they read,
\end{flushleft}
\onecolumngrid
%\begin{strip}
\bse
\bea
\alpha(t)=  %e^{-i \left(\zeta _k\right){}^-} 
\frac{g}{2 \epsilon  \left(4 \epsilon ^2-1\right)}\bigg[e^{i t \epsilon } \bigg(4 \epsilon ^2 \cos \left(t/2\right)-2 i \epsilon 
   \sin \left(t/2\right)-4 \epsilon ^2+1\bigg)-1 \bigg],
\ea
\bea
\theta(t) = -\frac{g^2}{2 \left(\epsilon -4 \epsilon ^3\right)^2}\Bigg[24 t \epsilon ^5-10 t \epsilon ^3+2 \left(4 \epsilon ^2-1\right) \epsilon ^3 \sin (t)+4 \epsilon ^2 \sin (t \epsilon
   )-4 \epsilon ^2 \cos \left(t/2\right) \sin (t \epsilon )\Bigg.\\
   \Bigg.-2 \epsilon  \sin \left(t/2\right) \bigg(1-\cos (t \epsilon
   )+32 \epsilon ^4-12 \epsilon ^2\bigg)+t \epsilon -\sin (t \epsilon )\Bigg]. \notag
\ea
\ese
%\end{strip}
%\twocolumngrid
For the $\sin(t/2)^2\cos(t)$ in an time interval $t\in [0,2\pi]$ they are
%\onecolumngrid
%\begin{strip}
\bse
\bea
\alpha(t)=\frac{ i g %e^{-i \left(\zeta _k^-+\zeta _k^+\right)}
}{8} \left(\frac{e^{i t (\epsilon -2)}-1}{\epsilon
   -2}-\frac{2 \left(e^{i t (\epsilon -1)}-1\right)}{\epsilon -1}+\frac{2 \left(e^{i t \epsilon }-1\right)}{\epsilon }-\frac{2\left(e^{i t (\epsilon +1)}-1\right)}{\epsilon +1}+\frac{e^{i t (\epsilon +2)}-1}{\epsilon +2}\right),
\ea
\bea
\theta(t)= \frac{g^2 }{192 \epsilon ^2 \left(\epsilon ^4-5 \epsilon ^2+4\right)^2}\left\{\epsilon  \left(\epsilon ^4-5 \epsilon ^2+4\right) \Bigg[12 t \left(7 \epsilon ^4-27 \epsilon ^2+8\right)+\epsilon ^2
   \bigg(8 \left(5-2 \epsilon ^2\right) \sin (3 t)+3 \left(\epsilon ^2-1\right) \sin (4 t)\bigg)\Bigg.\right.\notag\\
   \Bigg.\left.-24 \left(6 \epsilon ^4-23 \epsilon
   ^2+8\right) \sin (t)+24 \left(2 \epsilon ^4-7 \epsilon ^2+2\right) \sin (2 t)\Bigg]+96 \epsilon  \left(\epsilon ^2+2\right) \sin
   (t) \left(2 \left(\epsilon ^2-1\right) \cos (t)-\epsilon ^2+4\right) \cos (t \epsilon )\right.\\
   \left.-48 \left(\epsilon ^2+2\right) \sin (t
   \epsilon ) \Bigg(\left(\epsilon ^2-1\right) \left(\epsilon ^2 \cos (2 t)+\epsilon ^2-4\right)-2 \epsilon ^2 \left(\epsilon
   ^2-4\right) \cos (t)\Bigg)\right\}\notag.
\ea
\ese
%\end{strip}
%\twocolumngrid
Finally, for the $\sin(3t/2)^2$ modulation also for $t\in[0,2\pi]$ they read,
%\onecolumngrid
%\begin{strip}
\bse
\bea
\alpha(t)=\frac{g %e^{-i \zeta _k^-}% 
\Bigg(-2 \left(\epsilon ^2-9\right) e^{i t \epsilon }+(\epsilon -3) \epsilon  e^{i t
   (\epsilon +3)}+(\epsilon +3) \epsilon  e^{i t (\epsilon -3)}-18\Bigg)}{4 \epsilon  \left(\epsilon ^2-9\right)},
\ea
\bea
\theta(t)= \frac{g^2}{24 \epsilon ^2 \left(\epsilon ^2-9\right)^2}\left\{\left(9-\epsilon ^2\right) \left[\epsilon ^3 \sin (6 t)+4 \left(9-2 \epsilon ^2\right) \epsilon  \sin (3 t)+18 \bigg(t
   \epsilon  \left(\epsilon ^2-6\right)+6 \sin (t \epsilon )\bigg)\right]\right.\notag\\ \left.+54 (\epsilon -3) \epsilon  \sin \bigg(t (\epsilon +3)\bigg)-54
   \epsilon  (\epsilon +3) \sin \bigg( t(3- \epsilon ) \bigg)\right\}. 
\ea
\ese
%\end{strip}
In a similar fashion $\alpha$ and $\theta$ can be derived for the time intervals of the sequences in Section~\ref{section:three}.
\twocolumngrid

%-----------------------------------------

\end{document}